\documentclass[12pt]{article}
\usepackage[tmargin=1in,bmargin=1in,lmargin=1.25in,rmargin=1.25in]{geometry}

\usepackage{setspace}
\usepackage{csquotes}
\usepackage{amsmath}
\usepackage{amsfonts}
\usepackage{stmaryrd}
\usepackage{dsfont}
\usepackage{graphicx}
\graphicspath{ {images/} }

\begin{document}

	\title{Infinitesimal Gunk\footnote{I thank Jeffrey Russell for his very valuable input to multiple drafts of the paper. I thank Philip Bricker for his helpful feedback on early drafts of the paper. Thanks to Cian Dorr for his encouraging comments. Thanks to Tobias Fritz for helpful discussions. Special thanks to two anonymous referees of \textit{Journal of Philosophical Logic} for their scrupulous read, very helpful comments, and for pressing me on important details. }}
\author{Lu Chen}
	\date{(Forthcoming in \textit{Journal of Philosophical Logic})}
	\maketitle

\textbf{Abstract.} In this paper, I advance an original view of the structure of space called \textit{Infinitesimal Gunk}. This view says that every region of space can be further divided and some regions have infinitesimal size, where infinitesimals are understood in the framework of Robinson's (1966) nonstandard analysis. This view, I argue, provides a novel reply to the inconsistency arguments proposed by Arntzenius (2008) and Russell (2008), which have troubled a more familiar gunky approach. Moreover, it has important advantages over the alternative views these authors suggested. Unlike Arntzenius's proposal, it does not introduce regions with no interior. It also has a much richer measure theory than Russell's proposal and does not retreat to mere finite additivity.
\vspace*{4mm}

\vspace{10mm} 
	
	\section{Is Space Pointy?}
	
	Consider the space you occupy. Does it have ultimate parts? According to \textit{the standard view}, the answer is yes: space is composed of uncountably many unextended points.\footnote{ ``Space" can be understood as physical space or (mathematical) geometric space: the discussions in this paper do not turn on the differences between them. Many considerations also apply to time or spacetime.} Although standard, this view leads to many counterintuitive results. For example, intuitively, the size of a region should be the sum of the sizes of its disjoint parts.\footnote{This is one of the intuitions behind Zeno's paradox of measure. See Skyrms (1983) and Butterfield (2006).} But according to the standard view, the points have zero size. Thus they cannot add up to a finite size, because zeros always add up to zero. 
	
	For another example, according to the standard view, every region of space (except the whole space) has a boundary, and a closed region includes its boundary. Now, suppose that two rigid bodies which occupy closed regions come into perfect contact: there is no gap between them. Under the standard view, we cannot put two closed regions side by side without overlapping and without leaving a gap between them. Thus, to be in perfect contact, the two closed regions must overlap on their boundaries. But the bodies are rigid and impenetrable, so they should not occupy overlapping regions. Therefore, if the standard view is true, two rigid bodies that occupy closed regions cannot come into perfect contact. But perfect contact is intuitively possible. This is called ``the contact puzzle."(See Zimmerman 1996, Artnzenius 2008 and Russell 2008)
	
	Due to these problems, a \textit{gunky} conception of space has been proposed, according to which space cannot be broken down into ultimate parts. That is, every part of space can be further divided, and extensionless points do not exist. Such a conception can be traced back to the ancient Greeks, such as Anaxagoras.\footnote{``Nor of the small is there a smallest, but always a smaller..."(Curd 2007, B3)} Its contemporary development is often associated with A. N. Whitehead (1919, 1920, 1929).  Under Whitehead's theory, all regions have at least a finite size. So, it avoids the counterintuitive result that an extended region is composed of unextended points. Moreover, the contact puzzle can be avoided by denying the existence of boundaries. Call this approach \textit{the finite gunky view} (``finite" as opposed to the infinitesimal approach that I shall soon introduce). However, both F.\@ Arntzenius (2008) and J.\@ Russell (2008) pointed out that the finite gunky view, in conjunction with other plausible assumptions, is inconsistent with countable additivity, an attractive measure-theoretic principle. These authors proposed their own solutions, but at the expense of some attractive features of the original view. Arntzenius suggests readmitting boundaries with nonzero measures, even though they are scattered points with no interiors. Finding this proposal unattractive, Russell suggests rejecting countable additivity instead and having merely finite additivity. But the resulting measure theory is impoverished.

	However, the \textit{finite} gunky view is not the only approach to the gunky conception of space: a different approach is to claim that space also has parts of \textit{infinitesimal} sizes, which can be further divided. It has been argued that such a notion of divisible infinitesimals appeared in the Chrysippean doctrine of space, time and motion (White 1992). Although it has an ancient origin, such a doctrine was never developed further, due to the alleged obscurity of an infinitesimal size. But the situation has changed since the development of nonstandard analysis by Abraham Robinson (1966), which gives infinitesimals a rigorous foundation.\footnote{This is not the only foundation for infinitesimals. I explore alternative theories, such as \textit{smooth infinitesimal analysis}, in other work (Chen, manuscript). For my work on atomistic space in the framework of nonstandard analysis, see Chen (2019).} In this paper, I will develop a gunky view of space, \textit{Infinitesimal Gunk}, in the framework of nonstandard analysis. Like the finite gunky view, this view implies that every part of space can be further divided, and there are no boundaries. But unlike the finite gunky view, it implies that some parts have infinitesimal sizes. Developing such a view is not straightforward, for novel technical difficulties arise as we turn to nonstandard analysis. Thus part of my goal is to solve these difficulties and present a rigorous and most plausible gunky view in the framework of nonstandard analysis. In addition, I will advance Infinitesimal Gunk as a novel reply to the inconsistency arguments of Arntzenius and Russell. I will argue that Infinitesimal Gunk has distinctive advantages over the solutions proposed by these authors. Unlike Arntzenius's proposal, it does not need to admit regions without interiors. It also has a much richer measure theory than Russell's proposal. Infinitesimal Gunk also violates countable additivity, but it has attractive measure-theoretic compensations unavailable to Russell's proposal.

	\section{Trouble for the Finite Gunky View}

	Before I present \textit{the finite gunky view}, I shall first lay out the main ideas of a gunky space without assuming that every region of space has at least a  finite size. The intuitive ideas of a gunky space can be put into the following: 
	
	\begin{quote}
		\textit{Gunky Space.}\footnote{While the notion of gunky space is usually asssociated with only the mereological aspect, I am anticipating a more developed theory.} (Mereology) Every region has a proper part. (Topology) There are no boundary regions.  (Measure theory) Every region has a subregion of a strictly smaller size.

	\end{quote}

 \noindent The mereological aspect can be considered the starting point or the core claim of any gunky view of space. Since extensionless points have no proper parts, it follows that there are no points in space. The topological aspect needs some explanation. A boundary of a region is generally lower-dimensional than the region itself: it's like the skin of an apple if we idealize by imagining the skin to have no thickness at all.\footnote{The precise definition of ``boundary" in gunky space will be given later.} The requirement that there are no boundaries thus reflects the ``gunky" intuition that space has no lower-dimensional parts. This intuition is in a similar spirit to the mereological aspect: just as there are no indivisible points, there are no lines or surfaces in a higher-dimensional space because they cannot be divided along a particular dimension.\footnote{This reason does not apply to boundaries in general, since boundaries such as the fusion of two points in a one-dimensional space can be divided into two points.} Finally, the measure-theoretical aspect is also closely associated with the mereological aspect. It follows from the measure-theoretic principle that every region has a positive measure together with the mereological aspect and some plausible assumptions.\footnote{More explicitly, we need the following assumptions: (1) \textsc{Weak Supplementation}: if a region $x$ has a proper part $y$, then $x$ also has a proper part $z$ that is disjoint from $y$; (2) \textsc{Finite Additivity}: for finitely many regions, the size of their fusion is the sum of the sizes of those regions.} The principle that every region has a positive measure is motivated by the consideration that no extended region is entirely composed of unextended ones. 
	
	Before getting to the finite gunky view, we need to know a few topological terms. In standard topology, \textit{openness} is the only primitive topological notion: a \textit{topological space} is a set together with some choice of its open subsets (satisfying certain constraints). A \textit{closed} set is the complement of an open one. The \textit{interior} of a set is the union of its open subsets---like the flesh of an apple inside its ideally thin skin. The \textit{closure} of a set is the intersection of the closed sets including it---like a whole apple to its flesh. Further, a set that is identical to the closure of its interior is called \textit{regular closed}. For instance, consider the real line $\mathbb{R}$ (with its standard topology). The singleton of a point is not regular closed, because the closure of its interior is empty. Similarly, a set that includes an isolated point is not regular closed (e.g.\@ $[0,1]\cup\{2\}$). Now, every equivalence class of sets of real numbers that differ at most on their boundaries includes exactly one regular closed set. For the finite gunky view, the intuitive idea is that, since boundaries do not exist, every region should correspond to exactly one such equivalence class (except that of the empty set). In that case, every region can be represented by the regular closed set in its corresponding equivalence class. For simplicity, I will henceforth pretend that our space is one-dimensional (most discussion can be carried over to higher-dimensional cases straightforwardly). In the finite gunky view, we postulate the following principle for gunky space, which will be further strengthened later. 
	
	\begin{quote}
		\textsc{Real Representation}. There is a one-to-one correspondence between all regions of space and all non-empty regular closed sets of real numbers such that a region $X$ is a part of a region $Y$ iff $X$'s corresponding set is a subset of $Y$'s corresponding set.\footnote{Note that this principle (along with the measure specified later) implies that there are infinite regions of space. Although whether physical space is infinite or not need not to be settled by a gunky view, we postulate infinite regions for convenience. If one wish to have a gunky view of space with only finite regions, one can modify \textsc{Real Representation} easily. }
	\end{quote}

	\noindent As in standard mereology, other mereological notions can be defined in terms of parthood. For example, two regions \textit{overlap} iff they share a common part. Two regions are \textit{disjoint} iff they do not overlap. A region $X$ is a \textit{fusion} of regions $Y$s iff each $Y$ is a part of $X$ and every region that overlaps $X$ also overlaps one of $Y$s. For any collection of regions, their mereological fusion corresponds to the \textit{closure} of the union of their corresponding sets because the union of regular closed sets may not be regular closed and the closure of the union is the smallest regular closed set that includes those sets. For example, the fusion of the gunky regions represented by $[0,1/2],[0,3/4],[0,7/8]...$ is not represented by the union of those intervals, namely $[0,1)$, which is not regular closed, but by its closure $[0,1]$. Since every nonempty regular closed set includes a non-empty regular closed set as a proper subset, every region has a proper part---the mereological aspect of gunky space is confirmed. (For brevity, I will henceforth refer to nonempty sets by default unless otherwise specified.)

  We can also specify the topology of gunky space. In the standard topological framework, we have ``openness'' as the primitive notion. However, this framework is inadequate in the case of gunky space. Since we want there to be no boundaries, there should be no distinction between ``open" and ``closed" regions that differ at most on their boundaries.  As Roeper (1997) and other authors suggest, instead of ``openness,'' we can use the binary relation \textit{connectedness} as a primitive notion. To postulate the topology of gunky space, we strengthen \textsc{Real Representation} by the following clause:
   \begin{quote}
  A region $X$ is connected to a region $Y$ iff $X$'s corresponding set intersects $Y$'s corresponding set.  
   \end{quote}
   \noindent For example, $[0,1]$ and $[1,2]$ represent two connected regions since they intersect at 1. Following Russell (2008), other topological terms can be defined in terms of connectedness. For example, a region $X$ is a \textit{boundary} of a region $Y$ iff every part of $X$ is connected to both $Y$ and some region disjoint from $Y$. Intuitively, an apple's (ideally thin) skin is the boundary of the apple because every part of the skin is in contact with both the apple and its surrounding air. It follows that no region is a boundary of any region---thus the topological aspect of gunky space is met (Russell 2008, 7). Then we can define ``openness": a region is \textit{open} iff it does not overlap any of its boundaries.  It follows that every region is open. I will henceforth call  topologically strengthened \textsc{Real Representation}  together with the measure-theoretic principle that every region has a strictly smaller subregion \textit{the finite gunky view}.

	How should we measure regions? In standard analysis, the Lebesgue measure, which takes value in nonnegative extended real numbers $[0,+\infty]$,  is the standard way of assigning length, area, volume, and so on to subsets of a real coordinate space. Given \textsc{Real Representation}, it is natural to assign to a region the Lebesgue measure of its corresponding set. More precisely, we strengthen \textsc{Real Representation} by the following clause:
	
	\begin{quote}
		\textsc{Lebesgue Gunky Measure.} The measure of any region is equal to the Lebesgue measure of its corresponding set.
	\end{quote}
	
\noindent We can check that every region indeed has a subregion of a strictly smaller size. However, Arntzenius (2008) shows that this measure is not countably additive.

	\begin{quote}
	\textsc{Countable Additivity.}	For any countably many disjoint regions, their fusion has a measure, which is the sum of the measures of those regions.\footnote{Standard measure theory satisfies \textsc{Countable Additivity}. In particular, the fusion of countably infinitely many disjoint regions with positive measures has a measure of $+\infty$.}
\end{quote}

	\noindent The inconsistency argument runs as follows.  Consider a unit closed interval of real numbers. Take the middle closed interval of length 1/4. Then take the middle intervals of length $1/16$ from both of the remaining intervals. Repeat the same process \textit{ad infinitum}. In the $i$-th step, each middle interval we take is $1/4^{i}$-inch long. Label those intervals  $A_1, B_1, B_2, C_1,...$ such that the same letters refer to intervals of the same lengths. The intervals taken in the first three steps are illustrated below :
	
	\vspace{2mm}\includegraphics[scale=0.4]{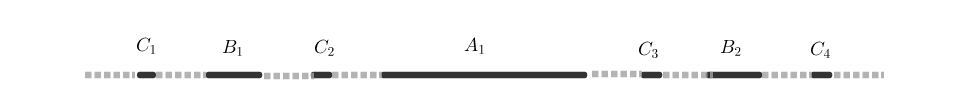}
	
	\noindent  Call the intervals \textit{the Cantor intervals} and their union \textit{the Cantor union}. A point is a \textit{limit point} of a set iff any open set including that point intersects that set. The closure of a set is the union of this set and all its limit points. We have that every point in the unit interval either belongs to the Cantor union or is a limit point of it. Thus, the closure of the Cantor union is the whole unit interval. Now, each Cantor interval represents a gunky region. Call those regions \textit{the Cantor regions}. It follows from \textsc{Real Representation} that the fusion of the Cantor regions is represented by the closure of the Cantor union, which is the whole unit interval. Thus, the Lebesgue gunky measure of the fusion is one. However, the Cantor regions that compose the fusion respectively measure 1/4, $1/16\cdot 2$, $1/64 \cdot 4$,..., which sum up to $1/2$. Suppose the measure theory is countably additive. Then the measure of the fusion is $1/2$. It follows that the measure of the fusion of the Cantor regions is both $1$ and $1/2$. So, \textsc{Lebesgue Gunky Measure} is not countably additive.
	
	What's worse, Russell (2008) pointed out that \textsc{Lebesgue Gunky Measure} is even inconsistent with finite additivity.  Let \textit{Big Fusion} be the fusion of the Cantor regions represented by \textit{Big Intervals} $A_1, C_1, C_2, C_3, C_4,...$, and let \textit{Small Fusion} be the fusion of the Cantor regions represented by \textit{Small Intervals} $B_1, B_2, D_1, D_2,..$.\footnote{The symbols are used as before: those with the same letter refer to intervals of the same length. Each of the $B$s is one-fourth the length of $A_1$, and each of the $C$s is one-fourth the length of each of the $B$s, etc.} It turns out that every point on the unit interval is either a point in the Cantor union or a limit point of \textit{both} the union of Big Intervals \textit{and} the union of Small Intervals. Then, the Lebesgue gunky measure of Big Fusion is the Lebesgue measure of the closure of the union of Big Intervals, which is 5/6. The measure of Small Fusion is the Lebesgue measure of the closure of the union of Small Intervals, which is 2/3. But the fusion of Big Fusion and Small Fusion is represented by the whole unit interval which measures one. Since Big Fusion and Small Fusion are disjoint regions, and $5/6+2/3\neq 1$, finite additivity fails!

	Russell gave a more general argument for the inconsistency between the finite gunky view,  \textsc{Countable Additivity}, and other plausible assumptions, which does not rely on \textsc{Lebesgue Gunky Measure} or any other specific measure. Instead, he appealed to an additional topological feature. In both standard topology and gunky topology, a \textit{basis} for a topological space is a set of \textit{basis elements} which generate all the open regions of the space by taking unions. A topological space can have many different bases. For instance, the set of all open cubes is a basis for the standard topology of three-dimensional real space $\mathbb{R}^3$. This implies that, for example, an apple-shaped open set in $\mathbb{R}^3$ is the union of many cubes. In standard topology, a topological space typically has a countable basis, that is, a basis with only countably many basis elements. For instance, one basis for the standard topology of the real line is the set of all open intervals with rational numbers as endpoints, which are countably many. Russell argued that, like a real coordinate space, our space has a countable basis.
	
	\begin{quote}
		\textsc{Countable Basis}. The topology of space has a countable basis.
	\end{quote}

	\noindent Russell has shown that the finite gunky view, if it satisfies \textsc{Countable Basis} and some standard mereological assumptions, is inconsistent with \textsc{Countable Additivity} (Russell 2008, 9).\footnote{The main mereological assumption is \textsc{Remainder Closure}, the principle that unless a region $X$ is proper part of a region $Y$, $X$ has a part that is the remainder of $Y$ in $X$ (or the mereological difference $X-Y$). (Russell 2008, 4)}

	In the face of the two inconsistency arguments, various solutions have been proposed. Arntzenius suggested denying that there are no boundary regions. Instead of \textsc{Real Representation}, Arntzenius suggested that a region should be represented by an equivalence class of Borel sets (formed from open sets through countable union, countable intersection and complement) that differ up to Lebesgue measure zero. Under his proposal, the fusion of the Cantor regions is distinct and smaller than the unit line segment, and there exists a boundary region that makes up the difference between the two regions. This region is represented by the equivalence class of the complement of the Cantor union and has no interior. Since this feature seems undesirable, we should check if there is a better alternative.\footnote{In this paper, I do not discuss why this feature of Arntzenius's theory is undesirable. I simply assume that it is intuitively attractive for a gunky space to have no boundaries (see Russell 2008).} Russell suggested that we should deny \textsc{Countable Additivity} and retreat to \textit{merely} finitely additive measure. Under this suggestion, the measure of any region is equal to the Jordan measure of its corresponding set (a restriction of the Lebesgue measure that is merely finitely additive). But the resulting measure theory is rather impoverished (which I discuss in Section 5). Is there another way out? Infinitesimal Gunk, the non-standard theory of space that I shall develop in Section 3, will provide a novel reply to the inconsistency arguments. In particular, \textsc{Countable Basis} fails, but without the costs that Russell assumed (Section 4). Although this theory also violates \textsc{Countable Additivity}, it satisfies a weaker version of it and has a much richer measure theory than Russell's proposal (Section 5). The theory is also immune to the variants of Arntzenius's argument (Section 6).

	\section{Infinitesimal Gunk}

	The core idea of Infinitesimal Gunk is that, instead of representing a gunky region by a set of real numbers, we represent a region by an extended set of numbers provided by nonstandard analysis (NSA). In NSA, we can extend the real line $\mathbb{R}$ to the hyperreal line $*\mathbb{R}$, which includes infinitesimals and infinite numbers, along with the familiar real numbers.\footnote{The hyperreal system I refer to in this paper is obtained through the ultrapower construction of real number sequences (see Goldblatt 1998). It's unique up to isomorphism under the assumption of continuum hypothesis (thanks to a referee for pointing this out). There are other non-isomorphic hyperreal systems. For example, the surreal number system introduced by Conway (1976) is considered the largest hyperreal system.} A number is \textit{infinitesimal} iff its absolute value is smaller than any positive real number. A number is \textit{infinite} iff its absolute value is larger than any positive real number. On the hyperreal line, each real number is surrounded by a ``cloud" of hyperreal numbers that are infinitesimally close to it, called its \textit{monad}. The monads of different real numbers do not overlap. Moreover, we can do arithmetic on the hyperreal numbers just like on the standard real numbers. For example, for any positive infinitesimal $\delta$  and any infinite number $N$, we have $\sqrt{\delta}$, $\frac{1}{\delta}$, $\delta^2$, $\delta+N$, $\frac{\delta}{N}$, etc. In general, the hyperreal numbers satisfy all the first-order truths about the real numbers in standard analysis. 
	
	As in the finite gunky view, I would like to represent gunky regions by regular closed sets. But unlike in the finite gunky view, I will appeal to sets of hyperreal numbers rather than just sets of real numbers. Before presenting the view, I shall first introduce the interval topology of the hyperreal line. (For a comparison of different topologies on the hyperreal line, see Goldblatt 1998, 120-1, 143-5.) According to the interval topology, a hyperreal set is \textit{open} iff it is the union of hyperreal intervals $(a,b)$ (i.e., the set of hyperreal numbers strictly between hyperreal numbers $a$ and $b$). For example, the set of all infinitesimals is open, because it is the union of all open intervals with infinitesimals as endpoints. Also, the set of all non-infinitesimals is also open, because it is the union of all open intervals of the form $(-N,-r)$ or $(r,N)$, with $r$ a positive real number and $N$ a positive infinite number. Since the complement of an open set is closed, the set of all infinitesimals is both open and closed, and so is the set of all non-infinitesimals. Furthermore, both sets are regular, since they are identical to the closure of their interiors. 
	
	Now, for Infinitesimal Gunk, I propose that regions are represented by regular closed sets of hyperreal numbers under the interval topology.
	
	\begin{quote}
	\textsc{Hyperreal Representation}. There is a one-to-one correspondence between all regions of space and all non-empty regular closed sets of hyperreal numbers such that a region $X$ is a part of a region $Y$ iff $X$'s corresponding set is a subset of $Y$'s corresponding set.\footnote{As in the finite gunky view, the set of all gunky regions forms a boolean algebra without the bottom element.} 
	\end{quote}
	
	\noindent As a basic example, the hyperreal interval $[0,1]$ is regular closed and thus represents a region. The set of all infinitesimals---and in general each monad---is regular closed, and thus represents a region. Furthermore, the countable union of $[1/2,1],[1/4,1],...$, $[1/2^n,1],...$ is also regular closed and represents a region.
	
We need to define a topology for gunky space so that there are indeed no boundary regions. Like in the finite gunky view, I will follow Roeper (1997) in using \textit{connectedness} as the primitive notion.  But the difficulty here is that we cannot postulate the connectedness relation between gunky regions in Infinitesimal Gunk in the same way as in the case of the finite gunky view. Recall that, in the finite gunky view, we postulate that two regions are connected iff their representative sets have a non-empty intersection. But this will not do the trick in Infinitesimal Gunk, because every monad represents a region, and distinct monads have no elements in common. If we postulate connectedness in the same way, then every region represented by a monad would be disconnected from the region represented by its complement on the hyperreal line.\footnote{Indeed, the interval topology itself has this problem: under the definition of connectedness in standard topology, every monad is disconnected from the rest of the hyperreal line.} This feature would be bad for our theory---for we want to describe a continuous space, which is composed of regions connected with each other. 

To solve this ``disconnection" problem, I propose an alternative to Roeper's postulation of connectedness between gunky regions. The intuitive idea is that two regions are connected iff they each contain a part such that there is no ``gap" between them. For any two sets $A, B$, let ``$A\leq B$" mean that for any $x\in A$ and any $y\in B$, we have $x\leq y$. Similarly, for any set $A$ and any point $z$, let ``$A\leq z$," for example, mean that for any $x\in A$ we have $x\leq z$. Then, the relation of connectedness satisfies the following principle:
	
	\begin{quote}
		\textsc{Connectedness.} Two regions represented by sets $A$ and $B$ are \textit{connected} iff there is a subset $A'$ of $A$ and a subset $B'$ of $B$ such that either (1) $A'\leq B'$, and there is no point $z$ with $A'<z<B'$, or (2) $B'\leq A'$, and there is no point $z$ with $B'<z<A'$.\footnote{In higher-dimensional space---informally speaking---two regions are connected iff they each contain a part such that there is no hyperreal hypersurface between their corresponding sets.} (Call such a $z$ a \textit{separating point}.)
	\end{quote}
	
	\noindent It immediately follows from this definition that a region represented by a monad is indeed connected with the region represented by the complement of the monad on the hyperreal line. Under this definition, connectedness is \textit{reflexive}, \textit{symmetric}, and \textit{monotonic}---the three features that constitute what Russel calls ``core topology" (Russell 2016, 262). In addition, connectedness is \textit{distributive}. Together, these features are essentially what Roeper took to be the ``central characteristics" of connectedness (Roeper 1997, 255).\footnote{A small complication is that Roeper's core axioms for connectedness assume the existence of the null region, while I assume there is no null region, which would require some small changes in the formalism. 
	
It is worth noting that my stipulation of connectedness does not satisfy all of Roeper's axioms beyond the core axioms. In particular, with any reasonable definition of another primitive notion \textit{limitedness} in Roeper's axioms, Infinitesimal Gunk would violate the following axiom (a region $X$ is \textit{well inside} a region $Y$ iff $X$ is not connected to $Y$'s complement):

\begin{quote}
	\textsc{A10.} If $A$ is a limited region, $B$ is not the null region, and $A$ is well inside $B$, then there is a limited region $C$ such that $A$ is well inside $C$ and $C$ is well inside $B$. (Roeper 1997, 256)
\end{quote}
In whatever way we define \textit{limitedness}, it is reasonable to assume that at least the hyperreal interval $[0,1]$ is limited. In the following sketch of a counterexample to A10, I will use Roeper's axiom that every part of a limited region is limited. A10 can fail for some infinite fusion of infinitesimal intervals. Let $\epsilon$ be an infinitesimal. Let $A$ be represented by the union of all intervals $[4n\epsilon,(4n+1)\epsilon]$ for all hypernatural $n$ such that $n\epsilon$ is an infinitesimal. Let $B$ be represented by the union of slightly larger intervals $[(4n-1)\epsilon,(4n+2)\epsilon]$ (with the same restriction on $n$) together with the set of all non-infinitesimal numbers. Note that $A$ is entirely contained in the monad of zero and thus limited, and it is well inside $B$. But there is no region $C$ that satisfies A10. Call a region \textit{snuggly} iff its representing set contains arbitrarily small positive non-infinitesimal numbers. It can be straightforwardly checked that, for any region $C$, if $A$ is well inside $C$, then $C$ is snuggly. But if $C$ is snuggly, then $C$ is connected with $B$'s complement. So A10 must be violated. 

One main role of Roeper's axioms is to ensure that there is a one-to-one correspondence between gunky topologies (or ``region-based topologies" in Roeper's term) and locally compact Hausdorff spaces under standard point-set topology. As a result of the violation of A10, we cannot recover a locally compact Hausdorff space from Infinitesimal Gunk through Roeper's correspondence (see Roeper 1997, 276, 278-9). This is not terribly surprising because the interval topology of the hyperreal line is not locally compact.

 }  Let $X,Y, Z$ range over all regions:
	
	\begin{quote}
		\textsc{Reflexivity.} $X$ is connected to itself.
		
		\textsc{Symmetry.} If $X$ is connected to $Y$, then $Y$ is connected to $X$. 
		
		\textsc{Monotonicity.} If $X$ is connected to $Y$, and $Y$ is a part of $Z$, then $X$ is connected to $Z$.
		
		\textsc{Distributivity.} If $X$ is connected to the fusion of $Y$ and $Z$, then $X$ is either connected to $Y$ or to $Z$.
	\end{quote}
I shall explain why \textsc{Distributivity} holds, since it's relatively less obvious. Take two arbitrary regions represented by sets $B$ and $C$. First, we note that the union of any two regular closed sets is still regular closed. Thus, the fusion of the two regions is represented by the union of $B$ and $C$. Suppose a region represented by set $A$ is connected to the fusion of those two regions. According to \textsc{Connectedness}, there are subsets $A'\subseteq A, B'\subseteq B$ and $C'\subseteq C$ with $A'\leq B'\cup C'$ or $B'\cup C'\leq A'$ such that there are no separating points between $A'$ and $B'\cup C'$.\footnote{Note that $B'$ or $C'$ in question could be empty, though they can't both be---every point separates the empty set from other sets.} It follows that either there are no separating points between $A'$ and $B'$ or there are no separating points between $A'$ and $C'$, for otherwise at least one of them would separate $A'$ and $B'\cup C'$. For example, if $A'\leq B'\cup C'$, then whichever of a separating point between $A'$ and $B'$ and one between $A'$ and $C'$ is smaller, it would separate $A'$ and $B'\cup C'$. So $A$ is either connected to  $B$ or to $C$.

	As in the finite gunky view, other topological terms are defined in terms of connectedness. 
	It follows from \textsc{Connectedness} that no region is a boundary of any region. Recall that a region $X$ is a boundary of a region $Y$ iff every part of $X$ is connected to both $Y$ and some region disjoint from $Y$. Suppose there is such a boundary region. Then it is represented by some regular closed set. The intuitive idea is that a regular closed set is ``fat" enough that we can always find a regular closed set strictly inside it. This smaller regular closed set represents a region that is disconnected from any region disjoint from the boundary region. This contradicts the definition of boundary, so there are no boundaries. It follows that the condition for openness is trivially satisfied, which means that every region is open. Also, every region is closed because every region has a complement (which is open). The closure of a region is always itself.

	Next, I shall postulate a measure over regions. Like in \textsc{Lebesgue Gunky Measure}, I will equate the measure of a region to the measure of its representing set. But first of all I shall propose a measure on the hyperreal line. The measure will be non-standard in the following senses. Instead of assigning nonnegative extended real numbers to subsets of a space, it assigns nonnegative \textit{hyperreal} measures to certain hyperreal sets. Also, unlike standard measure theory, the measurable sets are not closed under countable union, which I will discuss more in Section 5.\footnote{The measure I introduce here is similar to the ``proto-measure" introduced in Goldblatt (1998, 207). However, Goldblatt did not consider it a measure precisely because the measurable sets are not closed under countable union. He instead used it to define an extended-real-valued measure \textit{Loeb measure} which satisfies this requirement.}

	Similar to the construction of the Lebesgue measure on the real line, we first define the measure of a hyperreal interval:
	
	\begin{quote}
		\textsc{Interval Length.} For a hyperreal interval with end points $a,b$, its measure is $|b-a|$.
		
	\end{quote}
	
	\noindent Notice that the measure of an interval can be infinitesimal.
	
	Next, we define the length of the hyperreal set that is a union of such intervals.  But before that, I shall first explain the notion of \textit{hyperfinite sum} in NSA. First, notice the following fact: for any countably infinitely many items, even if they have an infinite sum (i.e., the limit of partial sums) in standard analysis, they generally do not have an infinite sum in NSA, because the partial sums do not converge to a unique hyperreal number. For instance, the partial sums $1/2+1/4+...+1/2^n$ do not converge to any unique hyperreal number.\footnote{There are two ways of defining ``converge" here. First, we can say that the partial sums $1/2+1/4+...+1/2^n$ \textit{converge} to a hyperreal number $h$, if their difference can be made smaller than any particular real number by making $n$ sufficiently large. Second, we can define ``converge" in a non-standard way: the partial sums $1/2+1/4+...+1/2^n$ \textit{converge} to a hyperreal number $h$, if their difference can be made smaller than any particular \textit{hyperreal} number by making $n$ sufficiently large. Under the first definition, the partial sums in question converge to many different hyperreal numbers. Under the second definition, the partial sums in question do not converge to any hyperreal number. Either way, there is no unique hyperreal number that the partial sums converge to.} In general, unlike the real line, the hyperreal line does not have the least upper bound property: the set of all the partial sums $1/2+1/4+...+1/2^n$ does not have a least upper bound. For every infinitesimal $\epsilon$, $1-\epsilon$ is an upper bound of the set, and there is no largest infinitesimal.
	
	However, there is a special kind of infinite ``cardinality," and accordingly a special kind of infinite sum in NSA. Recall that the set of all hyperreal numbers is an extension of the set of all real numbers. In the same sense, the set of all natural numbers can be extended to the set of \textit{hypernatural numbers}, which obey the same first-order truths of standard analysis as the natural numbers. Just as any real number is smaller than some natural number, any hyperreal number is smaller than some hypernatural number. Since there are infinite hyperreal numbers, it follows that there are also infinite hypernatural numbers. Let $N$ be a hypernatural number. In NSA, there is a distinct notion of ``cardinality"---call it \textit{hyperfinite cardinality}---that assigns $N$ to $\{1,2,...,N\}$, just as the finite set $\{1,2,...,n\} (n\in\mathbb{N})$ has a cardinality of $n$. A hyperfinite set is either finite or else continuum-sized. Furthermore, just as in standard analysis the sum of a finite sequence of real numbers is well-defined, in NSA, the sum of a hyperfinite sequence of hyperreal numbers is well-defined. This is called \textit{the hyperfinite sum.} Note though, hyperfiniteness is different from finiteness when it comes to higher-order claims: for example, a subset of a hyperfinite set need not be hyperfinite.\footnote{A subset of a hyperfinite set can be countably infinite, but no countably infinite set is hyperfinite.}  (For more on ``hyperfinite cardinality" and ``hyperfinite sum," see Goldblatt 1998, 178-81.)
	
	For any hyperreal set, if we can list the disjoint intervals it includes in a hyperfinite sequence, then its measure is the hyperfinite sum of the measures of those intervals.\footnote{A \textit{hyperfinite sequence} is an internal bijection from $\{1,2,...,N\}$, for some hypernatural $N$. An \textit{internal} function is a function that is expressible in the language of standard analysis. (Appendix A; see also Goldblatt 1998, 172-5 for more detail.)} 
	
	\begin{quote}
		\textsc{Hyperreal Measure.} For any hyperreal set, if it is a union of hyperfinitely many disjoint hyperreal intervals, then its measure is the sum of the measures of those intervals. Otherwise, its measure is undefined.
	\end{quote}

	\noindent Such a measure is well-defined because, like in the finite case, different decompositions of a hyperreal set into hyperfinitely many disjoint intervals (if possible) lead to the same measure.\footnote{That is, like in the finite case, it is true in nonstandard analysis that for any hyperfinitely many disjoint hyperreal intervals $B_1, B_2,...,B_N$ and $C_1, C_2,..., C_M$ ($N,M$ are hypernaturals), if the union of all $B_i$ is the same as the union of all $C_j$, then the hyperfinite sum of the measures of all $B_i$ is equal to the hyperfinite sum of the measures of all $C_j$. Note that the language of standard analysis quantifies over sets as well as numbers, and these quantifiers also receive nonstandard internal interpretation in the hyperreal system (Goldblatt 1998, 168-170).} Since every measurable set is a union of hyperfinitely many disjoint intervals, and because hyperfinite summation is associative like in the finite case, it follows that for hyperfinitely many disjoint measurable sets, the measure of their union is the sum of the measures of those sets.
	
	\begin{quote}
		\textsc{Hyperfinite Additivity (set)}. For hyperfinitely many disjoint measurable sets, the measure of their union is the sum of the measures of those sets.
	\end{quote} 
	
	The hyperreal measure approximates the Lebesgue measure over the real line in the following sense.  Recall that any finite hyperreal number is infinitely close to exactly one real number. The real number is called the \textit{shadow} of the hyperreal number. Let the shadow of any infinite positive hyperreal number be the extended real number $+\infty$. Let the shadow of a set be the set of the shadows of its members. Then, we have the following theorem:
	
	\begin{quote}
		\textsc{Lebesgue Approximation}. For any measurable hyperreal set, the shadow of its measure is the Lebesgue measure of its shadow (Goldblatt 1998, 215-7).
	\end{quote}

	\noindent In other words, the measure of a hyperreal set, if well-defined, is infinitesimally close to the Lebesgue measure of its shadow on the real line.

	Notice that some hyperreal sets, including some regular closed ones, are not measurable. For example, consider the set of all infinitesimals, which is regular closed. Intuitively, what should be the measure of such a set? The measure should be smaller than any real number. But it can't be an infinitesimal number because, for any positive infinitesimal number $\delta$, the set is larger than $(-\delta,\delta)$. Thus, the set of all infinitesimals has no measure. It's worth noting that some non-measurable hyperreal sets do have a Lebesgue-measurable shadow. For instance, the shadow of the set of all infinitesimals is \{0\} and therefore has Lebesgue measure zero.
	
	Finally, we can postulate the measure over a gunky line based on the measure over the hyperreal line and \textsc{Hyperreal Representation}:
	
	\begin{quote}
		\textsc{Hyperreal Gunky Measure.} The measure of a gunky region is the measure of its representing regular closed hyperreal set.
	\end{quote}
	
	\noindent For the reason I have just discussed, it follows that some gunky regions are not measurable.

	I have completed the basic picture of Infinitesimal Gunk, according to which every region is further dividable and some regions have infinitesimal sizes. Now, I shall evaluate this view in light of the inconsistency arguments and compare it with other solutions. 
	
	\section{No Countable Basis}
	
	Recall that Russell's inconsistency argument appeals to \textsc{Countable Basis}, the assumption that space has a countable basis. However, this does not hold for Infinitesimal Gunk. Informally speaking, infinitesimals are really small---in fact, for any merely \textit{countable} set of positive sizes, there are infinitesimal sizes which are even smaller than all of those. But that means that no countable collection of regions is ``fine-grained'' enough to build up all the regions. (See Appendix A.3 for my proof, which draws on a standard feature of the hyperreal system called \textit{countable saturation}.)
	
	When defending \textsc{Countable Basis}, Russell wrote,
	\begin{quote}
		There are topological spaces that do not have countable bases, but generally speaking, they are exotic infinite dimensional affairs. Such space would be shaped nothing like Euclidean space or any other ordinary manifold. (Russell 2008, 11)
	\end{quote}
	
	\noindent These claims are not quite justified in light of nonstandard analysis. The hyperreal line is indeed ``infinite dimensional" according to the standard definition of ``dimension": there is a homeomorphism from the hyperreal line to a Euclidean space of infinite dimensions.\footnote{A hyperreal number can be considered as an equivalence class of infinite sequences of real numbers that agree on ``almost" every position (which is defined through an ultrafilter). (Goldblatt, 1998)} But this definition is just inadequate to capture the geometric nature of the hyperreal line, namely that it's a one-dimensional line---one \textit{hyperreal} dimension.

	Denying \textsc{Countable Basis} in standard topology may result in ``exotic," ill-behaved spaces because it is associated with other desirable topological features of a space. For example, it is typically associated with \textit{metrizability}. A topological space is \textit{metrizable} iff we can define a real-valued distance between any two points such that the set of open balls with any radius are a basis for the topology. A \textit{metric space}, which is of special interest in physics and mathematics, is a metrizable space together with a specific distance function.  Every space with a countable basis is metrizable. Although the converse is not true in general, many commonly studied metrizable spaces have a countable basis.\footnote{Some rather unusual metric spaces do not have a countable basis. For example, consider any uncountable set. Let the distance between any two distinct elements be one. Then it generates a topology that does not have a countable basis (because for each element, its singleton is open). But such a distance function is not very interesting.}  So we typically require a space to have a countable basis to ensure that it is metrizable.
	
What hyperreal space shows is that it's not obvious that our physical space is metrizable, for it is completely natural to have a \textit{hyperreal-valued} distance function, rather than a real-valued one. Let's call the corresponding notion \textit{hypermetrizability}. Unlike metrizable spaces, it is not typical for a hypermetrizable space to have a countable basis---after all, a hyperreal space with the interval topology does not have a countable basis but is nevertheless hypermetrizable. For any two points on the hyperreal line $a, b$, we can define the distance between them to be $|b-a|$, which is the same as the length of the interval $(a,b)$. These open intervals constitute a basis for the interval topology of the hyperreal line.\footnote{Higher-dimensional cases are similar. For any two points in a hyperreal coordinate space $p=(p_1,p_2,...),q=(q_1,q_2,...)$, we can define the Euclidean distance between them, i.e., $d(p,q)=\sqrt{(p_1-q_1)^2+(p_2-q_2)^2+..}$. Then, all open balls constitute a basis for the interval topology of that hyperreal space.}

	\section{More Than Finite Additivity}
	
	I will now illustrate some desirable features of the measure in Infinitesimal Gunk by comparing it with the measure in Russell's solution. To avoid the inconsistency, Russell suggested rejecting \textsc{Countable Additivity} and using a merely finitely additive measure, such as the Jordan measure. Recall that in Arntzenius's inconsistency argument, \textsc{Countable Additivity} entails that the fusion of the Cantor regions has a measure of 1/2, but this fusion is represented by the unit interval, which has measure one. By rejecting \textsc{Countable Additivity}, Russell was able to claim that the fusion simply has measure one. To motivate this strategy, Russell argued that while finite additivity is necessary for understanding what a measure is, \textsc{Countable Additivity} need not be built into the nature of a measure. However, adopting a merely finitely additive measure has many drawbacks.
	
	To start with, it violates an attractive principle of supervenience: for countably many disjoint measurable regions, the measure of their fusion (or whether there is one) is completely determined by the measures of those regions. That is, for any countably many disjoint measurable regions,  no rearrangement will change the measure of their fusion (or whether their fusion has a measure). To put it more precisely:
	
	\begin{quote}
		\textsc{Countable Supervenience.} For any two countable sets $A,B$ of disjoint measurable regions, if there is a measure-preserving bijection from $A$ to $B$, then if the fusion of $A$ is measurable, the fusion of $B$ is measurable and has the same measure as the fusion of $A$.
	\end{quote}
	
	\noindent Russell's solution violates this principle. For instance, if you first walked 1/4 mile, and then 1/16 mile twice, and then 1/64 mile four times, and so on, in a straight line, then the total distance you walked is a half-mile. But when regions of these same sizes happen to be arranged like the Cantor intervals, the total distance becomes one mile. Thus, the measure of the fusion of countably many disjoint regions is not determined by the measures of those regions. This seems magical. Now, many people have argued that this violation is no more magical than the violation of supervenience in the uncountable case in the standard view, according to which the length of a line segment does not supervene on the length of its constituent points  (for example, see Hawthorne and Weatherson 2004). Thus, although the violation of \textsc{Countable Supervenience} may be technically inconvenient, it is philosophically no worse than the violation of uncountable supervenience, or the failure of arbitrary supervenience. But it is not obvious whether all motivations behind \textsc{Countable Supervenience} will generalize to all cases.\footnote{In probability theory, for example, Easwaran (2013) argued that some motivations behind countable additivity do not motivate uncountable additivity. So it is in principle possible to have considerations in favor of \textsc{Countable Supervenience} that does not generalize.} In standard mathematics, the countable case is usually more well-behaved than the uncountable case, so the countable case may be of special interest. In general, I will take it as an advantage to satisfy \textsc{Countable Supervenience} (which Infinitesimal Gunk does, as I shall argue soon), but I will leave it open whether this advantage is significant. 
	
	There is also the question what determines the measures of countable fusions. In many cases, no non-arbitrary answer can be given. Recall that Big Fusion, the fusion of the regions represented by $A_1, C_1,C_2,...$, and Small Fusion, the fusion of the regions represented by $B_1,B_2, D_1,...$, compose a region represented by the unit interval (see Section 2). But the Lebesgue gunky measures of  Big Fusion and Small Fusion are respectively 5/6 and 2/3. Since the resulting measure is not even finitely additive, Big Fusion and Small Fusion could not have these measures. What are their measures then? Russell suggested that we either consider them to be unmeasurable, or we assume brute facts about their measures that are compatible with finite additivity. We can assign to Big Fusion any value between its inner measure of 1/3 and its outer measure of 5/6, and to Small fusion any value between its inner measure of 1/6 and its outer measure of 2/3, as long as the sum of the two values add up to one (Russell 2008, 20-1). But these suggestions have clear drawbacks. If we consider those regions to be unmeasurable (without anything more to say), then the measure theory would be very restricted. But if those regions have measures, then the measure theory would involve many brute facts that are wildly different from their equally good alternatives.

	The situation for Infinitesimal Gunk is subtle. On the one hand, like Russell's solution, Infinitesimal Gunk also violates \textsc{Countable Additivity}. For example, suppose you are walking a straight line, and you first walked 1/2 mile, then 1/4 mile, then 1/8 mile, and so on for all natural numbers. How many miles have you walked in total? It's not one mile because the monad at the end of the one mile is not included in your journey. Indeed it's unmeasurable because the monad is not measurable. In fact, we can prove that for \textit{any} countably infinitely many disjoint measurable regions, their fusion is unmeasurable. It takes two steps to prove this claim. First, for any countably infinitely many disjoint measurable hyperreal sets, their union is unmeasurable (see Appendix A.5 for my proof based on countable saturation). Second, for any countably many disjoint regular closed sets, their union is regular closed (which, as I show in Appendix A.6, also follows from countable saturation). Since measurable regions are represented by measurable regular closed sets, it follows that for any countably infinitely many disjoint measurable regions, their fusion is always unmeasurable.

	 On the other hand, this very result entails that Infinitesimal Gunk satisfies \textsc{Countable Supervenience}, since the fusion of countably infinitely many disjoint measurable regions is always unmeasurable no matter how those regions are arranged. Thus, unlike in Russell's proposal, for countably many disjoint regions, the magic of changing the measure of their fusion (or whether there is a measure) through mere rearrangements of those regions does not occur.\footnote{Note that, like Russell's account or the standard measure theory, Infinitesimal Gunk does not satisfy arbitrary supervenience (or the restricted version of arbitrary supervenience that only involves \textit{internal} bijections) for reasons that will become apparent in the next section.} 
	
 However, one may argue that, even though \textsc{Countable Supervenience} is satisfied, the mere fact that \textit{no} fusion of countably infinitely many disjoint measurable regions has a measure is a serious cost to the theory. This may be true. However, what makes the measure theory of Infinitesimal Gunk attractive is that it has important compensations unavailable to Russell's proposal: the measure theory satisfies \textsc{Hyperfinite Additivity}, and we can define an extended-real-valued approximate measure that satisfies \textsc{Countable Additivity}. I will explain them in turn.

	 First, unlike Russell's solution, the measure theory does not retreat to mere finite additivity. Rather, it has hyperfinite additivity as a compensation. 
 
 	\begin{quote}
 	\textsc{Hyperfinite Additivity}. For hyperfinitely many disjoint measurable regions, their fusion has a measure, which is the (hyperfinite) sum of the measures of those regions.
 \end{quote}

 \noindent This follows from \textsc{Hyperfinite Additivity (Set)} in Section 3.\footnote{The principle can be reduced to \textsc{Hyperfinite Additivity (Set)} except for the caveat that two disjoint regions may be represented by two regular closed sets that overlap on their boundaries. But because every measurable set is the union of hyperfinitely many disjoint intervals, its boundary at most includes hyperfinitely many points. Since a point measures zero, and hyperfinitely many zeros add up to zero, the potential double counting of those boundary points in the case of overlapping boundaries does not affect the final measure.} As a result, Infinitesimal Gunk has a much richer measure than Russell's solution.
 
 To make this richness more vivid, consider \textit{*Big Fusion}, defined as the hyperreal extension of Big Fusion (that is, the fusion of the regions represented by the hyperreal extensions of $A_1, C_1, C_2,...$).\footnote{The hyperreal extension of $A_1$, for example, is the set of hyperreal numbers between the endpoints of $A_1$.} What is the measure of *Big Fusion? Infinitesimal Gunk implies that it is unmeasurable. But this is not all it says. The hyperreal measure satisfies the following theorem:
	
	\begin{quote}
		\textsc{Measure Approximation}. For any ``proper" hyperreal set $A$, if it has a Lebesgue measurable shadow, then there are measurable sets $B$ and $C$ such that $B\subseteq A\subseteq C$ and the measures of $B$ and $C$ differ by at most an infinitesimal.\footnote{The qualification of being ``proper" and having a Lebesgue measurable shadow corresponds to Loeb-measurablity (see Goldblatt 1998, 215-7). Then \textsc{Measure Approximation} follows from Goldblatt (1998, 212-4). Roughly, the set of Loeb-measurable sets is like the set of hyperreal-measurable sets except that it has more members so that it is closed under countable unions. The claims to follow in the main text hold under this qualification. In particular, the set representing *Big Fusion is Loeb-measurable.} (Goldblatt 1998, 212-4)
	\end{quote}
	
	\noindent Call a hyperreal set $C$ a \textit{Lebesgue completion} of a (proper) hyperreal set $A$ iff $A$ includes or is included in $C$ and $C$'s measure is equal to the Lebesgue measure of $A$'s shadow. Then, it follows that every (proper) unmeasurable set whose shadow is Lebesgue measurable has a Lebesgue completion.  Indeed, we can say that the hyperreal measure is richer than the Lebesgue measure in the following sense: there is an injective but not surjective function from Lebesgue-measurable sets on the real line to hyperreal-measurable sets on the hyperreal line that preserves their measures and mereological relations. In particular, if a Lebesgue-measurable set of real numbers is the shadow of an unmeasurable set, then this function takes the set of the real numbers to a Lebesgue completion of that unmeasurable set. In contrast, the Jordan measure is poorer than the Lebesgue measure in the same sense.
	
	Moreover, a natural notion of approximate measure can be defined based on \textsc{Measure Approximation}:
	
	\begin{quote}
		\textsc{Approximate Measure.} For any region, if its corresponding set has a Lebesgue completion, then it has an \textit{approximate measure}, which is equal to the measure of that completion. 
	\end{quote}
	Since *Big Fusion has a shadow that has a Lebesgue measure of 1/3, it has a Lebesgue completion that measures 1/3. This means that, even though *Big Fusion is unmeasurable, it has an approximate measure of 1/3. These approximate measures have nice properties. Notably, they satisfy \textsc{Countable Additivity}.
	
	\begin{quote}
		
		\textsc{Approximate Countable Additivity.} For any countably many disjoint regions that have approximate measures, their fusion also has an approximate measure, which is equal to the sum of the approximate measures of those regions.\footnote{The notion of approximate measure amounts to the Loeb measure, which is countably additive (Goldblatt 1998, 206-8; see also Footnote 32). Given that the Loeb measure is countably additive, it is natural to wonder why we do not use the Loeb measure instead of the hyperreal measure that I define in the paper. The main reason is that under the Loeb measure, all infinitesimal regions have zero measure. This violates the principle that every region has a strictly smaller subregion, which is one of the main intuitions behind the gunky approach to space. } (Goldblatt 1998, 206-8, 212-4)
		\end{quote}

	\noindent In comparison, in Russell's solution, no such attractive approximate measure can be systematically assigned. It is reasonable to assume that, under Russell's proposal, the approximate measure of a Jordan-measurable region is just its Jordan measure. Then, assuming 	\textsc{Approximate Countable Additivity}, the approximate measure of Big Fusion would again equal its inner measure $1/3$. Similarly, Small Fusion would have an approximate measure of $1/6$. But this would again violate even finite additivity, since Big Fusion and Small Fusion are disjoint regions, but their fusion has a measure of one.
	
	We began by looking for a theory of space that escaped Russell's theorem by giving up \textsc{Countable Basis}. What we have ended up with is a theory that gives up \textit{both} \textsc{Countable Basis} \textit{and} \textsc{Countable Additivity}. A point worth clearing up is whether this was inevitable. Is there a reasonable theory of space that gives up \textsc{Countable Basis} without also giving up \textsc{Countable Additivity}? Given some reasonable assumptions, the answer is no: there is no such theory.\footnote{
		Without any constraint, it is possible to violate \textsc{Countable Basis} without violating \textsc{Countable Additivity}, but this requires topological spaces that are too exotic to be a candidate for our actual physical space. A typical example is the product space $[0,1]^I$ with product topology, where $I$ is a cardinality larger than continuumly many. It does not have a countable basis, but has a countably additive measure. }
	 Here's a brief sketch as for why. If space does not have a countable basis, it would follow (under some reasonable assumptions) that there exist some very small regions that do not have any rational interval as a part. Moreover, assuming measure is translation-invariant, we can find countably infinitely many disjoint such regions with the same size within a unit interval. Those regions cannot all have positive finite measures. If they have infinitesimal measures (measures smaller than any finite number), then \textsc{Countable Additivity} would be violated. But if they all have measure zero, then the attractive measure-theoretic principle that all regions have smaller subregions would be violated.   Therefore, if we assume (among other things) that measure is translation-invariant and every region has a smaller subregion, it is impossible to give up \textsc{Countable Basis} without also giving up \textsc{Countable Additivity}.\footnote{I will demonstrate that it is impossible to violate \textsc{Countable Basis} without violating \textsc{Countable Additivity} under certain reasonable assumptions including that every region has a smaller subregion and measure is translation-invariant. As usual, I will pretend our space is ``one-dimensional'' in any suitable sense. The additional assumptions that the proof replies on are labeled in parentheses.
	 	
	 \textit{Proof}. Suppose space does not have a countable basis, and furthermore, the set of all measurable regions does not have a countable basis (although this supposition is stronger than the violation of \textsc{Countable Basis}, Russell's inconsistency proof effectively only involves the thesis that the set of all measurable regions has a countable basis). Also, suppose space is one-dimensional in the sense that its topology can be generated by some intervals (\textsc{Assumption-1}). In particular, we assume that every interval can be characterized by two endpoints, and that all endpoints are abstract entities that constitute a totally ordered Abelian group, to which rational numbers can be embedded. The set of all intervals with rational endpoints (or ``rational intervals'' for short) cannot be a basis since they are only countably many. As one can check, it follows that there is an interval $(\epsilon,\delta)$ that do not have any rational interval as a part and therefore $\delta-\epsilon$ is smaller than any rational number. Suppose measure is translation-invariant (\textsc{Assumption-2}). Then we can find countably infinitely many disjoint regions with the same measure as $(\epsilon, \delta)$ within the interval $(0,1)$. For example, let $\Delta=\delta-\epsilon$, and let $I_n=(\frac{1}{n}-\Delta, \frac{1}{n})$ for all $n\in \mathbb{N}$. Assuming \textsc{Countable Additivity}, the measure of the fusion of $I_n$ for all $n\in \mathbb{N}$ is well-defined---call this fusion ``Big.'' Now, consider the fusion of $I_n$ for all $n\geq 2$, and call this fusion ``Small.'' Then, given \textsc{Countable Supervenience} (which is entailed by \textsc{Countable Additivity}), Small has the same measure as Big. We further assume that for any two bounded regions, their measures have a well-defined subtraction (\textsc{Assumption-3}). Since Big is the fusion of the two disjoint regions $I_1$ and Small, it follows from finite additivity that $I_1$ has a measure of zero. This contradicts the principle that every region has a strictly smaller subregion (\textsc{Assumption-4}), which captures the measure-theoretic aspect of gunky space. Therefore, given the listed assumptions, it is impossible to violate \textsc{Countable Basis} without violating \textsc{Countable Additivity}. QED}

	\section{Variants of Arntzenius's Argument}

	Now let's examine how Infinitesimal Gunk avoids the specific difficulties raised by Arntzenius's Cantorian constructions as well as some variants. Arntzenius's original argument is based on \textsc{Real Representation}. Thus, it does not directly apply to Infinitesimal Gunk, since the regions are now no longer represented by regular closed sets of real numbers but of hyperreal numbers. So, let's consider the analogous argument on the hyperreal line. Take a unit hyperreal interval. Then similarly we cut out a 1/4-long hyperreal interval, two 1/16-long hyperreal intervals, four 1/64-long hyperreal intervals, and so on for all natural numbers---call them \textit{the *Cantor intervals} and their union simply \textit{*Cantor}. These intervals each represent a gunky region---call them \textit{the *Cantor regions}.
	
	Recall that Arntzenius's inconsistency argument relies on the following fact:
	
	\begin{quote}
		\textsc{Cantor Closure.} The closure of the Cantor union is the unit real interval. 
		
	\end{quote} 
	
	\noindent In the case of the hyperreal line, the analogous claim would be:
	
	\begin{quote}
		\textsc{*Cantor Closure}. The closure of *Cantor is the unit hyperreal interval. 
		
	\end{quote} 
	
	\noindent But this claim is false: the closure of *Cantor is not the whole unit line segment but rather *Cantor itself. Why? Consider a point in the unit interval that is not in any of the *Cantor intervals. One such point would be one that is infinitely close to the left endpoint of the unit interval. Let $\delta$ be an infinitesimal. $\delta$ is not in *Cantor since it is smaller than all the left endpoints of the *Cantor intervals. Furthermore, $\delta$ is not a limit point of *Cantor because there is an open set that includes $\delta$, but does not include any point in *Cantor. One example is $(0,\delta+\epsilon)$, where $\epsilon$ is any infinitesimal. Therefore, $\delta$ does not belong to the closure of *Cantor. In general, for any point $x$ outside *Cantor, $(x-\epsilon,x+\epsilon)$ is an open set that includes $x$ but does not overlap *Cantor. More vividly, each point outside *Cantor has an infinitesimal ``cushion" that ``protects" it from *Cantor. So, the closure of *Cantor on the hyperreal line is just itself.
	
	Can we come up with a different construction from *Cantor that gives rise to similar problems as in Arntzenius's argument?  One observation is that there is no analogous claim to \textsc{*Cantor Closure} as long as such a construction is composed of countably many intervals. 
	
	\begin{quote}
		\textsc{Countable Union}. The union of countably many disjoint closed intervals is regular closed.
	\end{quote}
	
	\noindent  Again, the idea is that infinitesimals are so small that for any point outside the countably many disjoint closed intervals, we can find an infinitesimally small neighborhood of that point disjoint from the union of those intervals. As a result, any point outside the union is not a limit point of it (Appendix A.6). 
	
	A perhaps cleverer revision of the argument is that, rather than cutting out countably many hyperreal intervals from the line segment, we cut out a much larger set of intervals. A countable set is in one-to-one correspondence with the set of all natural numbers. Analogously, let a \textit{hypercountable} set be in one-to-one (internal) correspondence with the set of all hypernatural numbers. The idea then is to cut out hypercountably many hyperreal intervals from the line segment: $R_1,R_2,...R_N,R_{N+1},...$, with $N$ being some infinite hypernatural. Call the union of these intervals \textit{Hypercantor}. In this case, every point outside Hypercantor is a limit point of it, because for any such point, its neighborhood---even if infinitesimally small---always intersects Hypercantor.\footnote{The reason for this is analogous to the reason why every point outside the Cantor union is a limit of the Cantor union. In the case of the Cantor union, for any point $x$ in the unit interval, for any positive real number $\epsilon$, we can find a point $y$ in the Cantor union such that $|y-x|<\epsilon$. Now, Hypercantor is constructed in the same way as the Cantor union, except that the cutting process does not stop with countably many intervals but continues for hypercountably many more. In particular, the Hypercantor intervals can be expressed in terms of when they are cut out in the same way as the Cantor intervals. For example, the leftmost point of the Hypercantor intervals cut out at stage $N$ is $1/2^{N+1}+1/2^{2N+1}$, just like the leftmost point of the Cantor intervals cut out at stage $n$ is $1/2^{n+1}+1/2^{2n+1}$. Thus, for instance, 0 is a limit point of Hypercantor because for any positive hyperreal $\delta$, we can find a hypernatural $N$ such that $1/2^{N+1}+1/2^{2N+1}<\delta$. This reasoning can be generalized to all the points on the unit hyperreal interval. } Consequently, the closure of Hypercantor  is the whole line segment! 
	
	But this does not cause trouble for Infinitesimal Gunk, because Hypercantor is not a union of hyperfinitely many hyperreal intervals.\footnote{There is no internal bijection between a hypercountable set and a hyperfinite set.} We can grant that the gunky region represented by the unit hyperreal interval is the fusion of hypercountably many gunky regions of lengths $1/4, 1/16,..., 1/4^N,...$. This does not result in any contradiction because we do not have ``hypercountable additivity" in Infinitesimal Gunk. (Notice that this reply has the same structure as Russell's reply to Arntzenius's original argument. That is, in the ``problematic" cases, both Russell and I rely on some version of additivity failing to hold. But as I discussed in Section 5, although hypercountable additivity fails, Infinitesimal Gunk still has several advantages over Russell's solution.)

	More generally, we can prove that any union of disjoint regular closed measurable sets that is not identical to its closure must be unmeasurable.
	
	\begin{quote}
		\textsc{Hyperfinite Union.} The union of hyperfinitely many disjoint measurable regular closed sets is regular closed.\footnote{This claim is an extension of the claim that the union of \textit{finitely} many regular closed sets is regular. See Appendix A.7 for my proof sketch.}
	\end{quote} 
	
	\noindent Moreover, any measurable regular closed set is the union of hyperfinitely many disjoint measurable regular closed sets (in particular, closed hyperreal intervals). It follows that any union of disjoint regular closed sets that is not regular closed is unmeasurable. As a result, Infinitesimal Gunk is safe from any Arntzenius-style trouble, since a union of regular closed sets that is not identical to their closures is unmeasurable and thus does not cause trouble, just like what we saw in the case of Hypercantor.

	\section{Conclusion}
	
	Can space be divided into ultimate parts? Does space have parts with infinitesimal sizes? These questions are more related to each other than they seem to be.   In this paper, I have shown that Infinitesimal Gunk, the view that any region of space can be further divided \textit{and} some regions are infinitesimally small, provides a novel reply to the inconsistency arguments given by Arntzenius and Russell. Moreover, this view has several important advantages over the solutions these authors suggested. It has a richer measure theory than Russell's proposal and satisfies attractive measure-theoretic principles unavailable to the latter. Unlike Arntzenius's proposal, it does not need to admit boundaries. Thus I recommend this novel theory for serious consideration.
	
	\newpage
	
	\appendix
	
	\noindent{\Huge Appendix}

	\section{Countable Saturation}
	
The language of standard analysis $\mathcal{L}$ that we focus on includes constant symbols for all real numbers and set constructions from real numbers through the iterations of the powerset operation and union.  More precisely, let $U_n(X)=U_{n-1}(X)\cup \mathcal{P}(U_{n-1}(X))$ and $U(X)=\bigcup_{n=0}^\infty U_n(X)$. The language includes constants for all members of $U(\mathbb{R})$, which have associated ranks according to the least $U_n(\mathbb{R})$ they belong. $\mathcal{L}$-terms and $\mathcal{L}$-formulas are defined in the usual way (Goldblatt 1998, 166-7).  We can show that all the familiar functions and relations in standard analysis (such as addition, integration, Lebesgue measure), considered as sets, are members of $U(\mathbb{R})$ and therefore are referred to by $\mathcal{L}$-constants (Goldblatt 1998, 165-6). Note that these functions and relations are not just defined over real numbers but can have a variety of ranks. The hyperreal system under consideration (along with the set constructions) $U'$ is an alternative model for $\mathcal{L}$ that is an expansion of $U(\mathbb{R})$. There is a unique transfer map from $U(\mathbb{R})$ to $U'$ that preserves all the $\mathcal{L}$-truths. Members of the image of the transfer map are called \textit{standard}. For example,  the hyperreal line $*\mathbb{R}$ is the image of $\mathbb{R}$ under the transfer map, and is therefore standard. An entity is \textit{internal} iff it is a member of a standard set (Goldblatt 1998, 172).  Any hyperreal number is internal because it is a member of $*\mathbb{R}$. Any hyperreal interval is internal because it is a member of the set $\{X\mid (\exists a,b\in*\mathbb{R})(\forall x\in X)(a\leq x\leq b)\}$, which is the image of the set of all real intervals. A bijective internal function belongs to a set of functions characterized as ``bijective" in $\mathcal{L}$ in the usual way. It turns out that all sets in the form of $\{x\mid \phi(x)\}$, where $\phi$ is a formula in the language $\mathcal{L}'$ that extends $\mathcal{L}$ with constants for internal entities, are internal sets (Goldblatt 1998, 177). For example, $\mathcal{L}'$ has constants for all hyperreal numbers. A hyperreal interval $\{x\mid a\leq x\leq b\}$ ($a,b\in *\mathbb{R})$ is therefore internal. On the other hand, we can prove that any infinite set of real numbers (e.g., $\mathbb{R},\mathbb{N}$) is not internal (Goldblatt 1998, 176).
		
	 In the hyperreal models we are interested in, the internal sets satisfy the following property:
	
	\newtheorem{saturation}[subsection]{Theorem}
	\begin{saturation}
		(Countable Saturation) The intersection of a decreasing sequence of nonempty internal sets $X^1\supseteq X^2\supseteq...$ is always nonempty (Goldblatt 1998, 138).
	\end{saturation}
	
	\noindent Countable Saturation implies this principle:
	
	\newtheorem{nested}[subsection]{Corollary}
	\begin{nested}
		(Nested Intervals) For any countable nested sequence of intervals $I_1\supseteq I_2\supseteq ...,$ their intersection is non-empty and includes an (open) interval.
	\end{nested}
	
	\noindent \textit{Proof.} All hyperreal intervals are internal sets. Thus, according to Countable Saturation, the countable nested sequence of intervals $I_1, I_2, ...$ have non-empty intersection. Moreover, the interiors of these intervals also have non-empty intersection. Let $x$ be a point in the intersection of their interiors. Then, the intersection of the parts of the intervals to the right of $x$ is non-empty. Let $y$ be a point in this intersection. Then, $[x,y]$ is included in the intersection of $I_1, I_2, ...$. (Clearly, the intersection also contains the open interval $(x,y)$.) QED.

With Nested Intervals, we can prove that Infinitesimal Gunk violates Countable Basis:

	\newtheorem{no countable basis}[subsection]{Theorem}
	
	\begin{no countable basis}
		Under Infinitesimal Gunk, the topology of space does not have a countable basis.
	\end{no countable basis}

\noindent\textit{Proof.} In this proof, we will use this fact: if a set of regions $\mathcal{B}$ is a basis for a gunky space, then every region in that space contains some region in $\mathcal{B}$. Let $\mathcal{C}$ be any countable set of regions. Take an arbitrary point $x$ on the hyperreal line, and consider the set of all elements in $\mathcal{C}$ that include $x$ in their interiors. Call this set $\mathcal{C}_x$. Since $\mathcal{C}$ is countable, $\mathcal{C}_x$ is also countable. It follows from Nested Intervals that there exists an infinitesimal neighborhood $\Delta$ of $x$ that is included in all elements of $\mathcal{C}_x$. Take a closed infinitesimal interval that is strictly included in $\Delta$. This interval does not contain any element of $\mathcal{C}$, so $\mathcal{C}$ is not a basis. Thus, a gunky space does not have a countable basis. QED.
	
	In Section 5 (p.23), I mentioned that the fusion of any countably infinitely many disjoint measurable regions is not measurable. This claim can be derived from the following corollaries of Countable Saturation:
	
	\newtheorem{countable}[subsection]{Corollary}
	
	\begin{countable}
		If an internal set $X$ is a countable union of internal sets $X_1, X_2,...$, then there is a natural number $k$ such that $X$ is the union of $X_1,...,X_k$ (Goldblatt 1998, 139).
	\end{countable}
	\noindent\textit{Proof.} The proof is adpated from Goldblatt (1998, 139-40). Suppose that for all $k\in\mathbb{N}$, $X- \bigcup_{n\leq k} X_n$ is non-empty. Since $X - \bigcup_{n\leq k}X_n = \bigcap_{n\leq k} (X-X_n)$, we have that $\bigcap_{n\leq k} (X-X_n)$ is non-empty. Call this set $Y^k$. Then $\langle Y^k\rangle$ is a decreasing sequence of non-empty internal sets. So, by Countable Saturation, there is a point belonging to $Y^k$ for all $k$, and thus to $X-X_k$ for all $k$. Therefore, $X$ is not the union of $X_1,X_2,...$. QED. 
	
	\newtheorem{fusion}[subsection]{Corollary}
	
	\begin{fusion}
		For any countably infinitely many disjoint measurable sets, their union is unmeasurable.
	\end{fusion}
	
	\noindent \textit{Proof}. Every measurable set is a union of hyperfinitely many disjoint intervals. This in fact guarantees that it is an internal set. Let $A_1, A_2,...$ be countably infinitely many disjoint measurable sets and let $A$ be their union. Suppose $A$ is measurable. According to Corollary B.4, it follows that $A$ is the union of finitely many $A_i$. But since $A_1,A_2,...$ are infinitely many and disjoint, their union is not identical to the union of any finitely many $A_i$. Thus, $A$ is not measurable. QED.

		\newtheorem{cu}[subsection]{Corollary}
		\begin{cu}
	 For any countably many disjoint regular closed sets, their union is regular closed.
	\end{cu}

	\noindent \textit{Proof.} Let $A_1, A_2,...$ be countably many disjoint regular closed sets on the hyperreal line. Let $A$ be their union. I will show that $A$ includes all its limit points and is therefore closed. Take any point $y$ outside $A$. For each regular closed set $A_j$, there is an open interval that includes $y$ and is disjoint from $A_j$. According to Nested Intervals, the intersection of these open intervals includes an open interval which includes $y$ and is disjoint from $A$. Thus $y$ is not a limit point of $A$.  Since $y$ is arbitrarily chosen, no point outside $A$ is a limit point. Therefore, $A$ is closed. QED.
	
\vspace*{4mm}

	In Section 6 (p.30), we need to show that there are no ``trouble-making" boundaries when it comes to the union of hyperfinitely many regular closed sets.
	
	\newtheorem{regular}[subsection]{Theorem}
	
	\begin{regular}
		For any hyperfinitely many measurable regular closed sets, their union is regular closed.
	\end{regular}

	\noindent \textit{Proof Sketch.} In standard analysis, we have the following induction principle for the natural numbers: an $\mathcal{L}$-formula $\phi$ with one free variable is satisfied by every natural number (taken as $1,2,...$) if (1) $\phi$ is satisfied by $n=1$; (2) if $\phi$ is satisfied by any natural number $n$, then it is also satisfied by $n+1$. In nonstandard analysis, we have an analogous induction principle for the hypernatural numbers:  an $\mathcal{L}$-formula $\psi$ with one free variable is satisfied by every hypernatural number if (1) $\psi$ is satisfied by $N=1$; (2) if $\psi$ is satisfied by any hypernatural number $N$, then it is also satisfied by $N+1$. Since any set with a hyperfinite cardinality $N$ can be ordered under an internal bijection to $\{1,2,...,N\}$, we will pick any such ordering of the set of hyperfinitely many measurable regular closed sets in question. Now, we can easily confirm the following: (1) the union of the singleton set of a measurable regular closed set is (trivially) regular closed; (2) if the union of the first $N$ measurable regular closed sets is regular closed, then the union of the first $N+1$ measurable regular closed sets is also regular closed because the union of two regular closed sets is regular closed. Also, these expressions can indeed be put into $\mathcal{L}$-formulas. Then according to the induction principle, for any hypernatural $N$, the union of the first $N$ measurable regular closed sets is regular closed, which is just what we want. QED.

\end{document}